# Quantum Software Engineering: A New Genre of Computing

Muhammad Azeem Akbar[1*], Arif Ali Khan[2], Sajjad Mahmood[3, 4], Saima Rafi[5]

[1]Department of Software Engineering, LUT University, Lappeenranta, Finland
[2]M3S Empirical Software Engineering Research Unit, University of Oulu, 90570 Oulu, Finland
[3]Information and Computer Science Department, King Fahd University of Petroleum and Minerals, Dhahran, Saudi Arabia.
[4]Interdisciplinary Research Center for Intelligent Secure Systems, King Fahd University of Petroleum and Minerals, Dhahran, Saudi Arabia.
[5]University of Murcia, Department of Informatics and Systems, Murcia, Spain
azeem.akbar@lut.fi, arif.khan@oulu.fi, smahmood@kfupm.edu.sa, saeem112@gmail.com
*corresponding author (Muhammad Azeem Akbar, azeem.akbar@lut.fi)

*Abstract*— Quantum computing (QC) is no longer only a scientific interest but is rapidly becoming an industrially available technology that can potentially tackle the limitations of classical computing. Over the last few years, major technology giants have invested in developing hardware and programming frameworks to develop quantum-specific applications. QC hardware technologies are gaining momentum, however, operationalizing the QC technologies trigger the need for software-intensive methodologies, techniques, processes, tools, roles, and responsibilities for developing industrial-centric quantum software applications. This paper presents the vision of the quantum software engineering (QSE) life cycle consisting of quantum requirements engineering, quantum software design, quantum software implementation, quantum software testing, and quantum software maintenance. This paper particularly calls for joint contributions of software engineering research and industrial community to present real-world solutions to support the entire quantum software development activities. The proposed vision facilitates the researchers and practitioners to propose new processes, reference architectures, novel tools, and practices to leverage quantum computers and develop emerging and next generations of quantum software.

*Keywords-Quantum computing (QC); Quantum software engineering (QSE); Quantum software development life cycle*

## I. Introduction

Quantum computing (QC) replaces the binary digits (bits) of classical electric computing with quantum bits (qubits), which, through features of quantum physics such as quantum states and quantum entanglement, will enable information to be processed exponentially faster than classical computers can manage [1, 2]. With quantum supremacy, in which a quantum computer can be shown to process any task faster than a classical computer on the horizon, researchers and companies are beginning to see some potential applications for exponentially faster computers than those in use today. These applications can improve our quality of life shortly. Machine learning powered by quantum computers promises to improve our quality of life, which is unimaginable [3].

QC promises to solve many problems more efficiently or precisely with classical computers, e.g., simulating complex physical systems or applying machine learning techniques [4, 5]. With recent advances in developing more powerful quantum computers, developing corresponding quantum software and applications and integrating them into existing software architectures are becoming increasingly important [6, 7]. However, the development of such quantum applications is complex and requires the knowledge of experts from various fields, e.g., physics, mathematics, and computer science [8].

Quantum software Engineering (QSE) is an emerging research area investigating concepts, principles, and guidelines to develop, maintain, and evolve quantum applications [8, 9]. It aims to increase the quality and reusability of the resulting quantum applications by systematically applying software engineering principles during all development phases, from the initial requirement analysis to the retirement of the software [10]. In classical software engineering, software development lifecycles (SDLC) are often used to document the different development phases a software artifact or application goes through [11, 12]. Furthermore, such SDLC also summarizes best practices and methods that can be applied in the various phases and corresponding tools [13-15]. Hence, they can educate new developers by providing an overview of the development process or serving as a basis for cooperating with experts from different fields [16, 17].

Building practical and real-life QC applications requires the implementation of quantum algorithms as software. Learning from the classical computing realm, developing dependable software entails following an SDLC, which typically includes requirements engineering, architecture and design, development, testing, debugging, and maintenance phases. Given that quantum software development is relatively new, a particular SDLC for quantum software does not exist. This paper aims to raise a voice for action concerning the importance of quantum software engineering in this era and to define a robust quantum software development life cycle.

## II. Background

Quantum software developers need novel techniques, tools, processes, and methods that explicitly focus on developing software systems based on quantum mechanics. Designing a quantum software algorithm is challenging because of fundamental quantum mechanics characteristics, including superposition and entanglement. New principles and methodologies for quantum software design are strongly demanded as the design is the most critical phase of developing the QS systems [3].

A. *Importance of Quantum in Emerging Technologies*

QC is an emerging area with significant potential, especially in optimization problems. Since QC works with a different mechanism than classical computing, the software approach for QC is also different. QC is primed to solve a broad spectrum of computationally expensive societal and industrial problems. Notable examples include accelerated drug discovery and vaccine development in healthcare, portfolio management, finance optimization, and complex physics simulations to understand our universe better. As a result, QC success will inevitably and significantly impact our day-to-day lives and revolutionize most industries across many domains. As examples, the implications of QC in emerging technologies are discussed below [18-20].

1) QC in Smart Cities

One application of a QC-based IoT would be a fully integrated and automated smart city of the future. This would manage energy production and distribution, waste treatment and disposal, pedestrian and vehicle traffic, lighting, and even atmospheric control [21]. Cities are proliferating with the increase in the human population, and cities of the future will need to accommodate more and more people to limit the effects of climate change on natural ecosystems. QC could enable the people living in cities to maintain a good quality of life despite the pressures of substantial local populations [22].

2) QC in Smart Road Networks

With smart cars and, in the near future, self-driving cars connected to a QC-based IoT, road accidents could be eradicated, fuel usage would be optimized for efficiency, and congestion could be significantly reduced [23, 24]. Thus, QC gives a potential solution for managing automatic driving in a vehicle in everything context.

3) QC in Smart Air Traffic Control

Similarly, air traffic control could be processed by quantum computers. This would deliver drastic improvements to accuracy (and, therefore, safety) and the manageable size of the ATC network [25]. This is necessary for a future where crewless aircraft (drones) will take over many of the physical delivery tasks currently performed inefficiently by humans driving on roads. It will enable quality-of-life improvements such as the quick, individual delivery of new products, medicines, and even passengers to destinations around large cities and beyond [26].

4) QC in Smart Factories

Suppose the speed of QC can be applied within the industrial automation sector. In that case, factories of the future will become vastly more efficient and able to take over more and more menial tasks currently completed or overseen by humans [27]. This would create economic benefits and free human labor to explore more meaningful ways of spending time [28].

5) QC in Smart Power Supply and Distribution

A QC-enabled electric grid linked to the IoT promises to remove inefficiencies from the power supply and distribution system, reducing human's need for energy while maintaining the modern quality of life [29]. QC-enabled modeling and forecasting. Classical computers simulating complex human and natural systems (like financial markets and the planet's climate) and using these simulations or models to predict the future accurately have already led to quality-of-life improvements in the twentieth century. This is known as 'forecasting. These can be significantly enhanced by introducing QC, gathering and taking inputs across entire complex systems, processing them, and predicting how they will interact [30].

6) QC-enabled Machine Learning

Machine learning, in which computers can learn and reprogram themselves for greater efficiency or accuracy in completing tasks, can bring unimaginable new applications for computers [31]. Enabled by the much more powerful future quantum computers, machine learning could advance rapidly. This, coupled with QC-enabled modeling and forecasting, has the potential to advance medicine and eradicate many diseases, provide treatment efficiently, and quickly diagnose and treat ailments (supported by the IoT) [31]. Further, machine learning enabled by quantum computers will improve our quality of life that cannot be imagined yet, as computers will continue to learn and develop themselves to support human goals of survival, harmony with the planet's ecosystems, and luxurious and effortless lives [32].

B. *Why quantum software Engineering*

Over the last few decades, QC has intrigued scientists, engineers, and the global public. Quantum computers use quantum superposition to perform many computations in parallel that are not possible with classical computers, resulting in tremendous computational power [33]. By exploiting such power, QC and quantum software enable applications typically out of the reach of classical computing, such as drug discovery and faster artificial intelligence (AI) techniques [34].

Quantum computers are currently being developed with various technologies, such as superconducting and ion trapping [34]. Private companies, such as Google and IBM, are building their quantum computers, while public entities invest in quantum technologies [8]. For example, the European Union Commission is spending €1 billion on quantum technologies ("EU's Quantum Flagship Project's Website"a). The key goal for quantum computers is to reduce hardware errors that limit their practical uses. Regardless of the eventual technology that wins the quantum hardware race, quantum software is the key enabler for building QC applications [8].

Quantum software needs to be supported with a quantum software stack, ranging from operating systems to compilers and programming languages [35]. Quantum computing's inherent characteristics, such as superposition and entanglement, and practical quantum software applications cannot be developed with classical software engineering methods. Moreover, software developers face significant challenges when coding quantum programs due to switching to an entirely different programming mindset with counter-intuitive quantum principles [14].

QSE needs to provide methods for developing quantum software. QSE requires tasks such as the design of quantum programs, implementation techniques for quantum algorithms, and testing and maintenance of quantum software [14]. Following the conventional wisdom in software programming, which started from hardware-focused, hard-wired techniques in the 1950s and then evolved into today's agile, iterative development, QSE should eventually become agile, iterative, and incremental [36]. Thus, we need to build novel QSE methodologies (with tool support) that cover different phases of QSE. Learning from the classical software engineering realm, developing trustworthy software entails following an SDLC, which typically includes requirements engineering, architecture and design, implementation, testing, and maintenance phases.

III. QUANTUM SOFTWARE DEVELOPMENT PHASES

QC is a technological revolution that demands a new software engineering paradigm to develop and conceive quantum software systems [14]. In QSE , developing a quantum software code is not the top priority problem, however, the requirements and design problems are much more common and important to correct [37]. Therefore, the focus on quantum software development techniques should not be limited to quantum coding issues but should overall focus on other aspects of QSE [38]. The promise quantum software development methodologies hold for overcoming complexity during analysis, design and accomplishing analysis and design reuse is significant [36]. If it is accepted that quantum software development is more than quantum programming, then a whole new approach, including life cycle, must be adopted to address other aspects of quantum software development.

QSE is the use of sound engineering principles for the development, operation, and maintenance of quantum software and the associated document to obtain economically quantum software that is reliable and works efficiently on quantum computers [35]. Thus, QSE calls for novel techniques, tools, processes, and methods that explicitly focus on developing software systems based on quantum mechanics. In the following sections, we briefly discuss the quantum software development life-cycle phases (i.e., quantum software requirements engineering, quantum software design, quantum software implementation, quantum software testing, and quantum software maintenance) [8, 14, 36, 39]. Table 1 presents the important artifacts of QSE life cycle, and a set of quantum software development activities derived from [8] and presented in Figure 1.

*A. Quantum Software Requirements Engineering*

We believe that quantum requirements engineering will be like requirements engineering process done for classical computing due to focus on requirement elicitation and management aspect of the phase. However, quantum requirements engineering will need new modeling and specification techniques to model QC aspects such as login functions and state. We believe that requirements engineering research community need to extend classical use cases, user stories and goal modeling techniques to support quantum requirements engineering process [14, 36].

*B. Quantum Software Design*

Like classical software development, quantum software design involves two major phases, i.e., architectural design and detail design. Architectural design is quantum software's abstract-level design, where the main components' interactions are described [39]. The detailed design explicitly describes the data structure, algorithms, and interfaces for module interactions. Developing algorithms for quantum software design is challenging compared to classical software systems because of some fundamental quantum computing features, including superposition and entanglement [8].

*C. Quantum Software Implementation*

Some work has been done to develop quantum programming languages [14]. There are several quantum programming languages, i.e., C (QCL), C++ (Scafflod), C\# (Q\#), Python (ProjectQ, Qiskit, Forest), F\# (LIQUi|)), Scala (Chisel-Q), and Haskell (Quiper) [14, 40]. However, research community need to focus on developing commercial integrated development environments to support quantum software implementation.

*D. Quantum Software Testing*

The testing phase begins to find defects and verify the system's behavior. It might be possible that the programmers make various mistakes when performing the quantum software testing because quantum computers have different properties such as superposition, entanglement, and no cloning, which make it challenging to predict the quantum software's behavior [14, 36]. There is a strong need for different testing tools and techniques that consider the quantum computing characteristics such as reading intermediate states [41], handling probabilistic test oracles and facing the decoherence problems.

*E. Quantum Software Maintenance*

The maintenance phase includes updating, changing, and modifying the quantum software to meet customer needs. This is an emerging research area in quantum software engineering, specifically focusing on re-engineering the available classical information systems and their integration with quantum algorithms[41]. In the short term, quantum computers might not be able to give full-fledged features because of the high cost and lack of processes, tools, and techniques [14, 36]. However, migrating the quantum algorithms to the existing classical information systems is the best option [42]. Therefore, it is important to revisit the reengineering (maintenance) concepts to deal with the migration process.

## IV. CALL FOR ACTION

Quantum computing is primed to solve a broad spectrum of computationally expensive societal and industrial problems. Notable examples include accelerated drug discovery and vaccine development in healthcare, portfolio management, finance optimization, and complex physics simulations to better understand the universe [43]. As a result, QC's success will inevitably and significantly impact our day-to-day lives and revolutionize most industries across different domains. Such impact must be realized via quantum software, the development of which should be systematically powered by quantum software engineering [14]. QSE opens new research areas to develop real applications by fostering research communities across disciplines (such as computer science, software engineering, mathematics, and physics) and interactions with other fields, such as medicine, chemistry, and finance. QC is on the rise and will revolutionize many areas of life. It will transform our understanding of and deal with complex problems and challenges. QSE is key to the systematic and cost-effective creation of tomorrow's robust, reliable, and practical QC applications [14, 42].

Despite recent advances in quantum programming tools, there are two significant challenges facing quantum software developers. First, there is an understandable fear of betting on a platform or language that ends up being discontinued. Second, quantum computer scientists must understand conventional software engineering principles, techniques and awareness of quantum programming [44]. Similarly, the complexity of writing quantum software is another challenge. Quantum programmers need to be experts in quantum information theory and have a working understanding of quantum physics and a mastery of mathematics [44, 45]. As a result, organizations face problems developing QSE teams [14, 36, 44, 45].

In a nutshell, we raise the call for action to develop the guidelines\frameworks for each phase of QSE life cycle, which could give a foundation for the develping a quantum software system. Furthermore, we argue that these phase-based guidelines\frameworks should not be an effort for a single researcher or research group, but a multidisciplinary project that builds on a combination of theoretical models and empirical results.

In the software development process, requirements engineering works as a foundation and steers quality software development as per expectations [37]. Quantum software requirements engineering requires a team aware of quantum mechanisms' computation and working style. Similarly, the requirements analysis team needs to develop quantum software engineering methods that specify the stakeholders' needs, quantum characteristics, and quantum attributes [14]. Features and properties of quantum, which are currently unclear, must be defined to perform functional and non-functional activities of a system. No guidelines on quantum mechanics are available to structure complex models. There is a need to develop new and effective hybrid (quantum-classic) guidelines to assist the requirement engineering to consider quantum mechanics while analyzing the requirement and developing the use case and associated UML models.

In QSE, the design phase gives a foundation to construct a hybrid structure of a software system. Hence, it is an essential step to model the quantum-classic characteristics of a software system. In hybrid software design, it is important to consider the system to be a set of components or modules with clearly defined behaviors & boundaries for each module [8]. A hybrid design should be correct, complete, efficient, flexible, consistent, and maintainable [8]. However, in this new genre of quantum software engineering, there is a need to build an understanding of architecting hybrid design software at both high- and low-level abstraction. To make the hybrid design efficient and complete, there is a need for new methods and tools to define abstract specifications, develop quantum modules, describe the functionality of each module, and understand the constraints and elements of quantum gates.

Table 1. Quantum software engineering artifacts

| | | | | | |
|---|---|---|---|---|---|
| **Quantum Software Requirements Engineering** | *Functional* | Define quantum features | Focus on user requirements with quantum | Capture quantum mechanism in the use case | Authentication and authorization levels |
| | *Non-Functional* | Define quantum properties | Focus on user expectations with quantum | Captured quantum as a quality attribute | Usability, reliability, scalability, performance |
| **Quantum Software Design** | *High-level design* | Quantum abstract specification | Develop Quantum Modules | Describe the functionality of each module | Understanding of quantum gates |
| | *Low-level design* | Quantum data structure design | Quantum Circuit and Algorithm design | Quantum interface design | Modules communication |
| **Quantum Software Implementation** | *Translating quantum circuit diagrams* | Specify Quantum programming language | Program; specifies the quantum circuit | Test using Quantum stimulator | Execute program on server |
| **Quantum Software Testing** | *Develop test plans* | Test design description | Test case description | Test reports and logs | Measuring the Qubits |
| **Quantum Software Maintenance** | *Determination of maintenance need* | Determination of maintenance scope | Quantum Architectural Design Maintenance | Quantum Modules Maintenance | Unit and acceptance testing |

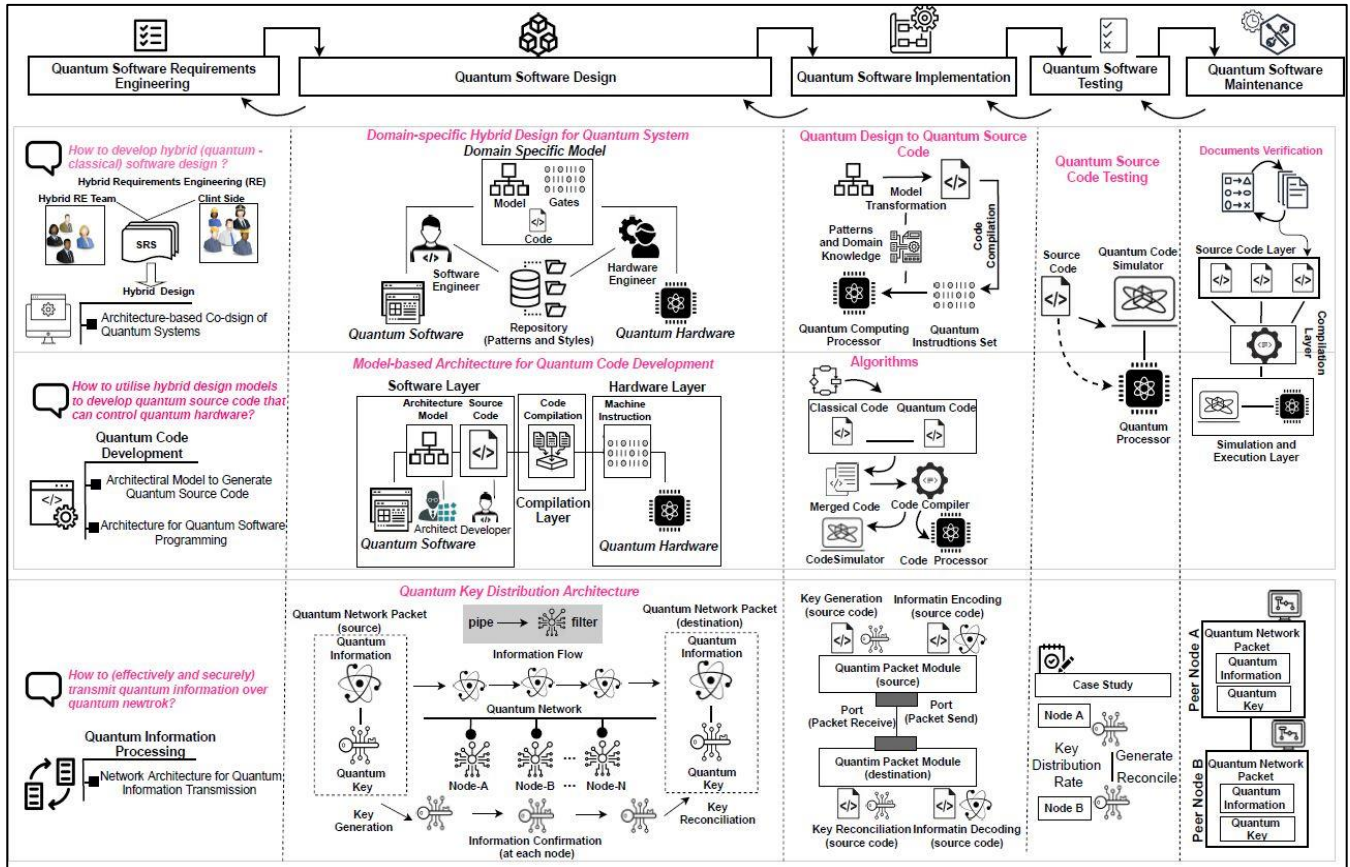

Figure 1. Quantum software development activities

Similarly, low-level abstraction, including data structure design, quantum circuit, algorithm design, quantum interface design, and module communications, are the areas that need the attention of practitioners and the research community to establish standard representations for quantum software components.

The implementation phase concerns software translating hybrid design specifications into the quantum source code. For quantum programming, the programmer requires a basic understanding of the quantum mechanism and software engineering [8, 14]. Using traditional software implementation methods is rare as we deal with circuits and quantum programming languages. We are in the early stage of quantum software implementation. We need more actions in terms of developing tools and technologies, quantum simulators, executing programs on servers, designing circuit diagrams, and developing of quantum programming roadmap. Implementing the code quality standards reduces many problems and the risk of project failures. Thus, there is a call for action for developing quantum programming standards and programming guidelines.

Software testing is very important in ensuring system quality control. It must ensure that the requested expectations of quantum software stakeholders have been satisfied [11]. The testing phase is also important to detect the errors of early phases that are related to project success as it directly reflects the project budget, time, and resources. As quantum software engineering is a new genre of software engineering, where quantum software has different properties such as superposition, entanglement, and no cloning, which makes it challenging to predict the behavior of quantum software [8, 14]. Dealing with quantum software testing using classical testing techniques and tools is not enough to make the quantum software error-free and up to the mark. This paper call for action to the development of testing tools, techniques, processes, and standards\guidelines considering the quantum characteristics of the software system.

Quantum software maintenance refers to the modifications created by correcting, inserting, deleting, extending, and enhancing the baseline of quantum software for successful maintenance, it is obligatory to estimate the demanded maintenance and the impact of change [36, 46]. There is a need for a quantum software experts-based change control board to identify, and verify the root cause, and effects of requested maintenance. As quantum software engineering is a new paradigm, there is a lack of quantum-oriented tools, techniques, and matrices to measure the scope of request maintenance. Hence, there is a demanding need for the develop of a set of tools, techniques, matrixes, and standards\guidelines for the quantum software maintenance process.

ACKNOWLEDGMENT

We are thankful to the research teams of LUT University, M3S Empirical Software Engineering Research Unit, University of Oulu, and King Fahd University of Petroleum and Minerals for giving their insights to come up with this vision paper. Moreover, this research is supported by the PHP Foundation with the grant 20220006.